\documentclass[useAMS,usenatbib]{mn2e}
\usepackage{graphicx}

\newcommand{\kms}{~km~s$^{-1}$}
\newcommand{\apj}{ApJ}

\def\av{A$_v$ }

\def\ergs{~erg~s$^{-1}$}

\def\cm3{~cm$^{-3}$ }
\def\msolr{~M$_\odot$~yr$^{-1}$}
\newcommand{\rate}{~cts~s$^{-1}$}

\title[]
{X-rays from Colliding Stellar Winds: the case of close WR$+$O 
binary systems}
\author[S.A.Zhekov]{Svetozar A. Zhekov
\thanks{E-mail: szhekov@space.bas.bg} \\
Space and Sollar-Terrestrial Research Institute, 6 Moskovska str.,
Sofia-1000, Bulgaria\\
}
\date{}

\pagerange{\pageref{firstpage}--\pageref{lastpage}} \pubyear{2012}

\begin{document}

\maketitle

\label{firstpage}

\begin{abstract}
We have analysed the X-ray emission from a sample of close WR$+$O
binaries using data from the public {\it Chandra} and {\it XMM-Newton} 
archives.  Global spectral fits show that two-temperature plasma is 
needed to match the X-ray emission from these objects as the hot 
component (kT~$> 2$~keV) is an important ingredient of the spectral 
models.  In  close WR$+$O binaries, X-rays likely originate in 
colliding stellar wind (CSW) shocks driven by the massive winds of the 
binary components. CSW shocks in these objects are expected to be 
radiative due to the high density of the plasma in the interaction 
region. Opposite to this, our analysis shows that the CSW shocks in 
the sample of close  WR$+$O binaries are {\it adiabatic}. This is 
possible only if the mass-loss rates of the stellar components in the 
binary are at least one order of magnitude smaller than the values
currently accepted. The most likely explanation for the X-ray
properties of close WR$+$O binaries could be that their winds are
two-component flows.  The more massive component (dense clumps) play 
role for the optical/UV emission from these objects, while the smooth
rarefied component is a key factor for their X-ray emission.

\end{abstract}


\section{Introduction}

As first proposed by \citet{pril_76} and \citet{cherep_76}, the
present concept on the origin of X-rays in massive Wolf-Rayet$+$O
(WR$+$O) binaries assumes that they arise in colliding-stellar-wind 
(CSW) shocks resulting from the interaction of the massive winds of 
the stars in the binary.
High wind velocities and high mass-loss rates of WR and O stars
suggest that the CSW region will be a luminous X-ray source. Thus,
WR$+$O binaries are expected to be brighter (more luminous) in X-rays 
than single WR  stars and, in fact, this was found in the early 
surveys of WR stars \citep{po_87}. 

It is interesting to note that wide WR$+$O binaries
are among the brightest WR objects in X-rays and their grating
spectra with the modern observatories ({\it Chandra, XMM-Newton}) show
numerous lines of various ionic species that indicate thermal emission 
(\citealt{sk_01}; 
\citealt{raa_03}; \citealt{schi_04}; \citealt{po_05}; 
\citealt{zhp_10b}). Analyses of these data confirmed that X-rays
from CSW shocks is the most probable physical picture for the X-ray
emission from wide WR$+$O binaries (although in some cases the observed
picture might be more complex than that; \citealt{zhp_10a},b).

On the other hand, it is noted from those early X-ray surveys
\citep{po_87} that 
close WR$+$O binaries are not as luminous in X-rays as the wide WR$+$O 
binaries. This is an important observation since the stellar wind
parameters (which in general determine the X-ray emission from CSWs)
are practically within the same range for both types of binary
systems. 

As a possible resolution to this problem, \citet{cherep_90}
proposed that most of the mass flux in the stellar wind
of a WR star is in the form of dense clouds. These clouds penetrate
freely the CSW region formed by the continuous component of the winds
and they do not contribute to the X-ray emission from the binary.
Thus, in a close WR$+$O binary only small part of the stellar wind gas 
(the continuous component) plays role for the X-ray emission from the
system. However, in a wide WR$+$O binary the CSW region forms far from
the stars where the dense clouds have already expanded and `dissolved'
in the continuous component of the wind, thus, the entire mass flux of
the stellar wind contributes to the X-ray emission from the binary.

Since the modern X-ray observatories ({\it Chandra, XMM-Newton})
provide us with data of much higher quality (both in sensitivity and
spatial resolution) compared to that of previous X-ray missions, we
decided to explore the case of CSWs in close WR$+$O binaries in some
detail. This was done for a number of such objects that have 
been observed by {\it Chandra} and/or {\it XMM-Newton}. We define the
sample of close WR$+$O binaries in Section~\ref{sec:thesample} and we
describe the corresponding X-ray data in Section~\ref{sec:data}. In
Section~\ref{sec:global_fits}, we present the results from the global
spectral models. In Section~\ref{sec:discussion}, we discuss our
results and we list our conclusions in Section~\ref{sec:conclusions}.

\section{The CSW Binaries Sample}
\label{sec:thesample}

In the sample of close CSW binaries, we included those objects of known
WR$+$O binaries (\citealt{vdh_01}; Tables 18 and 19) with orbital period 
smaller than 20-22 days. We have checked the archives of the modern X-ray 
observatories ({\it Chandra} and {\it XMM-Newton}) for available data on 
the thus defined sample. Some objects with X-ray detection are not 
included in our study and we particularly mention the following objects: 
WR48 (because it is a suggested triple stellar system and its X-ray
emission is quite complex; see \citealt{sugawara_08}); WR43a (since it is 
in the core of NGC 3603, thus, the object cannot be resolved); WR101k (it 
is in the center of the Galaxy and unresolved: being in $1\farcs8$ 
from Sgr A; SIMBAD).

Table 1 presents the list of close CSW binaries considered in this work 
and summarizes some of their properties.

\begin{table*}
\caption{Properties of the close WR$+$O binaries}
\label{tab:binaries}
\begin{tabular}{llllllll}
\hline
Name  & Spectral  &  \av  & Distance  & Period   & 
        V$_{\infty}$ &  $\dot{M}$  & Sources \\
      & type      & (mag) &  (kpc)    & (days)   &  (\kms) & (\msolr)
      & used for $\dot{M}$ \\
\hline
WR46  &  WN3$+$OB  &  1.05  & 4.07  & 0.329  & 2450 & 9.94e-06 &
 2, 3 \\
WR47  &  WN6$+$O5  &   3.96 &  3.80 &  6.24  & 1660  & 2.30e-05 & 
 6, 7, 10, 12 \\
WR139 &  WN5$+$O  &   2.83  & 1.90  & 4.21  & 1600 & 1.70e-05  &
 1, 4, 5, 8, 10, 11, 12, 13 \\
WR141 &  WN5$+$O5  &   4.10 &  1.26 & 21.7  & 1550 & 2.48e-05 & 
 1, 8, 10 \\
WR145 &  WN7/WCE$+$? & --- &  1.70  &  22.55  & 1390 & 3.68e-05 & 
 1, 10 \\
WR148 &  WN8$+$B3  &  2.58  & 8.28  & 4.32  & 1500 & 3.95e-05 & 
 8, 9, 12 \\
WR79  &  WC7$+$O5-8 &  1.54 &  1.99 &  8.89  & 2270  & 5.59e-05 & 
 1, 6, 10, 11, 12 \\
\hline
\end{tabular}

\vspace{0.5cm}
Note --  The object name, spectral type, interstellar extinction,
distance, orbital period and wind velocity are from the VII-th
catalogue of galactic Wolf-Rayet stars (\citealt{vdh_01};
$A_v = 1.11 A_V$) with the
following exceptions: (a) the wind velocity for WR47 is the mean value
for WN6 stars from \citet{eenens_94} and
for WR148 is from \citet{nishimaki_08}; (b) the orbital period for 
WR46 is from \citet{mar_00} and for WR145 is from \citet{muntean_09};
(c) the distance to WR145 is that to Cyg OB2.
The mass-loss rates are the mean values for each object as found in 
the literature with no clumping taken into account.  The individual
data were scaled correspondingly if derived at values for the distance 
to the object and/or its stellar wind velocity different from those 
adopted here.
The corresponding references are:
(1) \citet{ab_86}; 
(2) \citet{crowther_95};
(3) \citet{hamann_98};
(4) \citet{hirv_06}; 
(5) \citet{kurosawa_02};
(6) \citet{lamon_96}
(7) \citet{moffat_90};
(8) \citet{nishimaki_08};
(9) \citet{nugis_98}; 
(10) \citet{nugis_00}; 
(11) \citet{prinja_90};
(12) \citet{st_louis_88}; 
(13) \citet{st_louis_93}.
\end{table*}

\section{X-ray Data}
\label{sec:data}

Since the studied objects are not very bright in X-rays, we made use
of the available data from the Advanced Camera for Imaging Spectroscopy
(ACIS-S) on-board {\it Chandra} and of the data from the European Photon
Imaging Camera (EPIC; pn and MOS detectors) on-board {\it XMM-Newton}.

For the analysis of the ACIS-S data, we extracted the X-ray spectra 
following procedures in the Science Threads for Imaging Spectroscopy of 
the CIAO \footnote{Chandra Interactive Analysis of Observations 
(CIAO), http://cxc.harvard.edu/ciao/} 4.3 data analysis software.
The response functions and ancillary response functions for all spectra
were generated using the {\it Chandra} calibration data base CALDB v4.4.2.

For extracting X-ray spectra from the EPIC (pn, MOS) data, we made use
of the {\it XMM-Newton} SAS \footnote{Science Analysis Software (SAS), 
http://xmm.esac.esa.int/sas/} 10.0.0 data analysis software. 
The SAS pipeline processing scripts emproc and epproc were
executed to incorporate the most recent calibration files.
The data were then filtered for high X-ray background following the
instructions in the SAS documentation (e.g., adopting typical threshold 
rates for high-energy background in pn and MOS: 0.4 and 0.2 cts s$^{-1}$, 
respectively). The SAS 
procedures rmfgen and arfgen were adopted to generate the corresponding 
response matrix files and ancillary response files for each spectrum.
For each data set, the MOS spectra in this analysis are the sum of the
spectra from the two MOS detectors.

Table 2 presents the basic information about the X-ray data of the
objects in our sample of close CSW binaries. Also, some details on
the individual objects are given below.

\begin{table*}
\caption{X-ray Observations}
\label{tab:data}
\begin{tabular}{lllllll}
\hline
Name  &  Observation  &  Binary  &  Detector  &  Source  &   Rate   &  Data  \\
      &     IDs       &  phase   &            &  counts  &  (\rate) &        \\
\hline
WR46  &  0109110101   &  0.55    &    pn  &  3790 &   6.20e-02  & 3,
4, 5  \\
      &               &          &    MOS &  2605 &   3.62e-02  &   \\
WR47  &  0109480101   &  0.62    &    pn  &  1911 &   4.50e-02  & 1  \\
      &               &          &    MOS &  1665 &   3.22e-02  &   \\
      &  0109480401   &  0.89    &    pn  &  1004 &   2.86e-02  &   \\
      &               &          &    MOS &   843 &   1.99e-02  &   \\
WR139 &  0206240201   &  0.47    &    pn  &   1514 &  0.176  & 1, 2 \\
      &  0206240701   &  0.55    &    pn  &   1723 &  0.214  &   \\
      &  0206240801   &  0.81    &    pn  &   3744 &  0.273  &   \\
WR141 &  0404540201   &  0.02    &    MOS &   734 &   5.30e-02  & 5  \\
WR145 &  10969        &  0.33    &    ACIS-S &   757 &   2.64e-02  &   \\
      &  0165360101   &  0.00    &    MOS &      297 &   2.06e-02  &   \\
      &  0165360201   &  0.09    &    MOS &      366 &   2.54e-02  &   \\
      &  0165360401   &  0.18    &    MOS &      361 &   2.50e-02  &   \\
WR148 &  0405060201   &  0.99    &    pn  &  231 &   1.11e-02  &   \\
      &               &          &    MOS &  214 &   7.48e-03  &   \\
      &  0405060301   &  0.44    &    MOS &  107 &   8.32e-03  &   \\
WR79  &  5372         &  0.72    &    ACIS-S &   384 &   5.04e-03  &   \\
      &  6291         &  0.23    &    ACIS-S &   203 &   4.57e-03  &   \\
\hline
\end{tabular}

\vspace{0.5cm}
Note --  The binary phase for each observation was calculated using
the value for T$_0$ from Pourbaix et al. (2004; The ninth catalogue of
spectroscopic binary orbits) except for WR46 (see notes to
Table~\ref{tab:binaries}). The orbital `phase' for WR145 is 
with respect of the first in time X-ray observation since no value for 
T$_0$ is available in the literature. The moment T$_0$ (binary phase 
0.0) is the time of maximum value of the radial velocity of the WR 
component in the binary system. 
The values for the orbital period in \citet{poir_04}
are identical to those in Table~\ref{tab:binaries} except for 
WR46 and WR145. For these two, we used more recent sources given in the 
notes to Table~\ref{tab:binaries}.
The last column gives the references where the corresponding X-ray data
have been already discussed in some detail:
(1) \citet{bhatt_10}; (2) \citet{fauchez_11};
(3) - \citet{goss_11a}; 
(4) \citet{hen_br_11}; (5) \citet{naze_09}.
\end{table*}

{\bf   WR46. }
There is one pointed observation of this object with {\it XMM-Newton}.
The EPIC data are of good quality (no long background flares), thus, all
the spectra, pn and combined MOS, were included in this analysis.
This is the object with the shortest binary period in our sample (see
Table~\ref{tab:binaries}), thus, the EPIC spectra are average
(integrated) over the binary period (also, see Appendix~\ref{app} for 
discussion of the variability issue).

{\bf   WR47. }
There are three pointed observations of this star in the {\it
XMM-Newton} archive and only two of them (see Table~\ref{tab:data}) 
were used in this study since the third one showed considerable 
high-energy X-ray background rates. Both pn and MOS spectra were 
included in the current analysis.

{\bf   WR139. }
There are six pointed observations of WR139 carried on with {\it
XMM-Newton}, and here we used the pn spectra of three of them  that
have  good photon statistics. The basic reason for the lower quality
of the rest of the data sets is the high X-ray background which
shortens their effective exposure.

{\bf   WR141. }
There are two data sets in the {\it XMM-Newton} archive that detect
this WR binary at $\sim 5\farcm1$ from the optical axis.
Unfortunately, due to the high X-ray background rate, only part of the
MOS exposure in the second observation could be used in our analysis.

{\bf   WR145. }
We find one {\it Chandra} and three {\it XMM-Newton} archive
observations that have detected this WR star respectively at
$\sim5\farcm2$ and $\sim 9\farcm6$ from the optical axis.
The quality of the EPIC data is not high and only the MOS spectra
could be used in this analysis.

{\bf   WR148. }
In the {\it XMM-Newton} archive, there are two pointed observations of
this distant WR binary. Due to high X-ray background, the pn spectrum
of the second data set was excluded from our analysis.

{\bf   WR79. }
We find two {\it Chanrda} and eight {\it XMM-Newton} archive
observations respectively at $\sim 1\arcmin$ and $\sim 3\arcmin$ 
from the optical axis, but only the former provide useful X-ray spectra 
of this WR object. The reason is that there are two near-by 
sources (within $6\arcsec$ from WR79) that cannot be spatially 
resolved by {\it XMM-Newton} (see Appendix~\ref{app} for details).

\section{Global fits}
\label{sec:global_fits}

Since it is expected that the X-ray emission from the close WR+O
binaries originates in the interaction region of their massive winds,
a systematic approach to modelling these spectra could be to use a
hydrodynamic CSW model. However, the quality of the current data
(undispersed CCD spectra with not very good photon statistics for
most of the spectra) is not a good basis for carrying on such a
modelling in detail.
This is why, we decided to use conventional, discrete-temperature
plasma, models for the global fits to the X-ray spectra of the objects
in our sample. Thus, we adopted the optically thin plasma model
{\it apec} in  a recent version (11.3.2) of the XSPEC analysis package
\citep{Arnaud96}. Since the shocked WR gas dominates the X-ray
emission from CSWs in WR$+$O binaries (the mass loss of a WR star is
about an order of magnitude higher than of an O star),
in the global spectral fits we kept the abundances of the chemical
elements fixed to their values typical for the WN and WC
stars, correspondingly \citep{vdh_86}.
To derive the X-ray absorption towards each object, we used the
\citet{morrison_83} cross-sections (model {\it wabs} in XSPEC). And it
is worth noting that  we found X-ray absorption higher than
that expected from the optical/UV observations. Since the interaction
region in close CSW binaries should be located near the massive stars,
it is realistic to attribute this extra X-ray absorption to
the massive stellar winds of these objects. For sake of simplicity,
we assumed that the extra absorption is due to
the stellar wind of the WR star in the system because it is
considerably more massive than that of its companion.

Thus, the models for our final global spectral fits consisted of the
following components. An absorption component due to the interstellar
matter towards the studied object (model {\it wabs} in XSPEC). The
value of the neutral hydrogen column density  was kept fixed to that
based on the optical/UV observations (e.g. \av in
Table~\ref{tab:binaries}) through the \citet{go_75} conversion
(N$_H = 2.22\times10^{21}$A$_V$~cm$^{-2}$). Anticipating the
discussion of the derived results, we note that our conclusions from
this study do not change if we use a more recent conversion
(\citealt{vuong_03}; \citealt{getman_05}; N$_H =
1.6-1.7\times10^{21}$A$_V$~cm$^{-2}$).
Emission components are
represented with discrete-temperature optically-thin plasma  models
({\it apec} in XSPEC) with individual `wind' absorption for each 
plasma component. The emission and `wind' absorption components had 
the same abundances.
We recall that in the framework of the CSW picture, the
discrete-temperature models represent qualitatively the temperature
stratification of the interaction region (see \S 5.2 in
\citealt{zhek_07}).
We note that there are no objects in our sample of close CSW binaries
that have elliptical orbits. This means that the amount of hot gas in
the interaction region remains the same over an orbital period.
So, for each object that had multi-phase observations,
we adopted the same model normalization for the corresponding plasma
component. This is equivalent to a non-variable emission measure with
the orbital phase that is there is no variation in the intrinsic X-ray
emission. We thus attribute a possible X-ray variability to the
variable X-ray absorption with the orbital phase.

One-temperature model was successful for fitting the X-ray spectrum of
WR141. We found that two-temperature models gave adequate global fits
for the X-ray emission from all other WN and WC stars in the sample
but WR46. For the latter, we needed a three-temperature  model. We
note that the two-temperature one gave a statistically acceptable result
but it was not able to match all the spectral details in the spectrum:
e.g., it produced almost none emission for the Si XIII line complex at
$\sim 1.85$~keV and variable silicon abundances could not fix this
problem. It is our understanding that a more complex model was
needed in this case due to the good photon statistics of the WR46 spectra
(the best quality data in our sample). Also, we had no data on the
interstellar absorption towards WR145, we thus fitted for the X-ray
absorption in the global spectral fits for this object.

\begin{figure*}
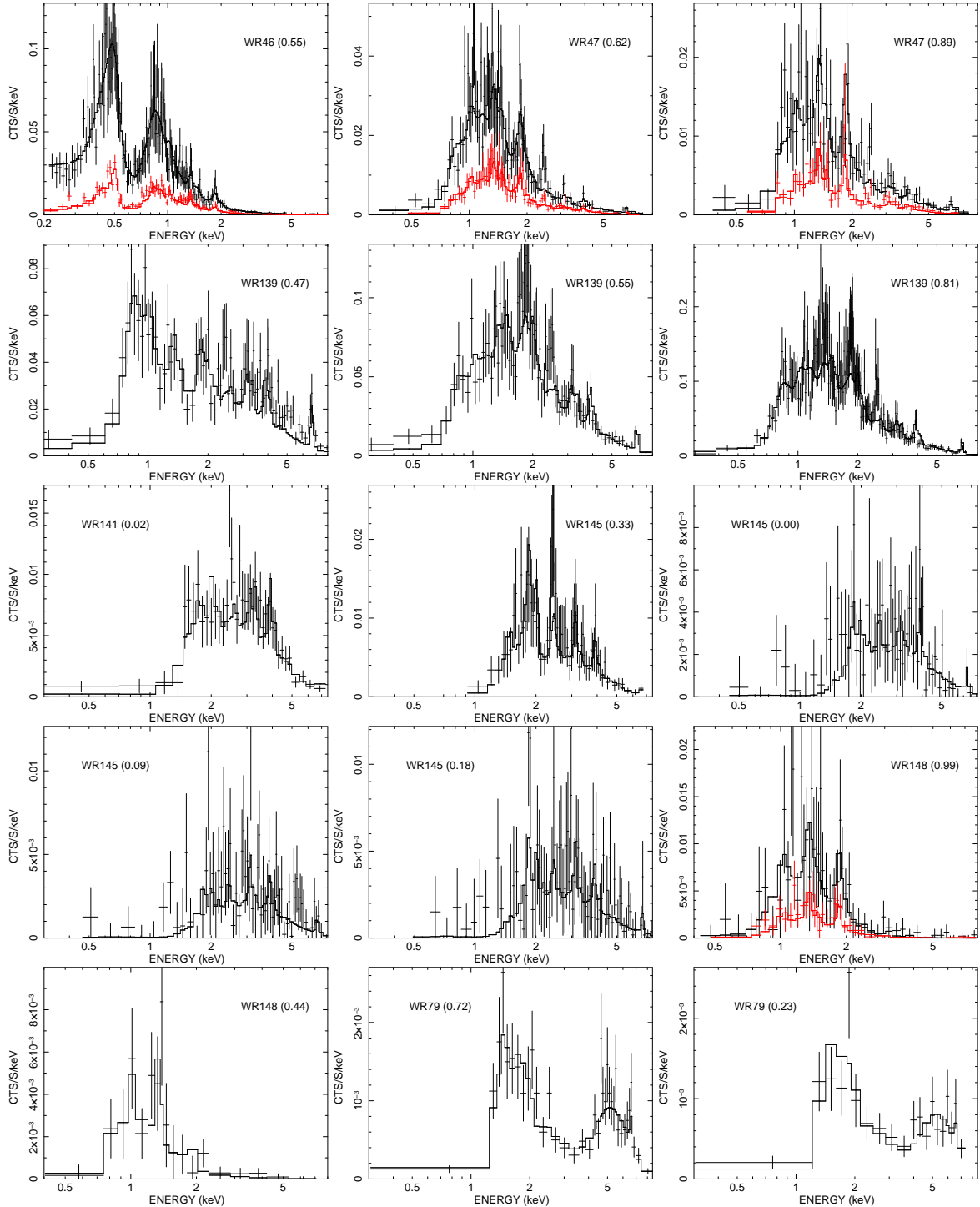

 \centering\includegraphics[width=1.5in,height=2.in,angle=-90]{fig1a.eps}
 \centering\includegraphics[width=1.5in,height=2.in,angle=-90]{fig1b.eps}
 \centering\includegraphics[width=1.5in,height=2.in,angle=-90]{fig1c.eps}
 \centering\includegraphics[width=1.5in,height=2.in,angle=-90]{fig1d.eps}
 \centering\includegraphics[width=1.5in,height=2.in,angle=-90]{fig1e.eps}
 \centering\includegraphics[width=1.5in,height=2.in,angle=-90]{fig1f.eps}
 \centering\includegraphics[width=1.5in,height=2.in,angle=-90]{fig1g.eps}
 \centering\includegraphics[width=1.5in,height=2.in,angle=-90]{fig1h.eps}
 \centering\includegraphics[width=1.5in,height=2.in,angle=-90]{fig1i.eps}
 \centering\includegraphics[width=1.5in,height=2.in,angle=-90]{fig1j.eps}
 \centering\includegraphics[width=1.5in,height=2.in,angle=-90]{fig1k.eps}
 \centering\includegraphics[width=1.5in,height=2.in,angle=-90]{fig1l.eps}
 \centering\includegraphics[width=1.5in,height=2.in,angle=-90]{fig1m.eps}
 \centering\includegraphics[width=1.5in,height=2.in,angle=-90]{fig1n.eps}
 \centering\includegraphics[width=1.5in,height=2.in,angle=-90]{fig1o.eps}
\caption{
The background-subtracted spectra  overlaid with the best-fitting
model.
Each panel is marked by the name of the object and the binary phase in
parentheses (see Tables~\ref{tab:data} and ~\ref{tab:fits}).
The spectra are re-binned to have a minimum of 10-20 counts per bin.
In the panels with two different spectra in the same data set, the
upper and lower (in red) curves are for the pn and MOS spectrum, 
respectively.
}
\label{fig:spectra}
\end{figure*}

Figure~\ref{fig:spectra} and Table~\ref{tab:fits} present the results
from the global fits to the X-ray spectra of our sample of close WR$+$O
binaries. We see that the quality of the fits is acceptable but we note
that in many of the studied cases the formal quality of the fit
($\chi^2$/dof) is good likely due to the low photon statistics.
Nevertheless, we note that the fit results firmly establish that a 
considerable amount of hot-temperature plasma (kT~$> 2$~keV) is present 
in the studied objects. 
The `canonical' WN and WC abundances adopted here 
\citep{vdh_86}  seem to be an adequate approximation to
the actual abundances in the studied objects. Although, a by-product
from these fits in this respect is worth mentioning. Namely, as seen
from Fig.~\ref{fig:spectra} the SXV line complex at $\sim 2.45$~keV is
not well matched by the current model. Our attempts to solve this
problem by adding an additional temperature component were not  
successful. The only way to match the SXV `bump' in the spectra was to
adopt a variable sulfur abundance. Interestingly, the values for
the S abundance derived from all fits are not widely spread but
'cluster' around a factor of three increase with respect to the one in
the `canonical' WN and WC abundances. More specifically, the average
sulfur abundance from all data sets is  $2.97 \pm 0.33$ (where the
error is the error of the mean and not the variance). We note that the
sulfur value in the \citet{vdh_86} WN and WC abundances
is based on the cosmic mass fraction (see the footnote to their Table
1), thus, we get an indication from the current analysis that the
sulfur abundance in the `canonical' WN and WC sets of abundances may
need be increased by a factor of three.

In the next section, we will discuss the global fit results in the
framework of the CSW picture.

\begin{table*}
\caption{Global Spectral Model Results}
\label{tab:fits}
\begin{tabular}{lllllllllll}
\hline
Name & Phase$^{(a)}$ & $\chi^2$/dof & N$_{H,~ISM}$$^{(b)}$  &
       N$_{He,~1}$$^{(c)}$ & kT$_1$$^{(d)}$ & EM$_1$$^{(e)}$ &
       N$_{He,~2}$$^{(c)}$ & kT$_2$$^{(d)}$ & EM$_2$$^{(e)}$ &
       F$_X$$^{(f)}$ \\
\hline
WR46  &       0.55    & 231/255  & 2.10  &
         
$0.002^{+0.34}_{-0.002}$  & $0.15^{+0.01}_{-0.01}$  &
         $5.55^{+31.5}_{-0.27}$  &
         $0.35^{+0.09}_{-0.09}$  & $0.57^{+0.02}_{-0.02}$  &
         $3.75^{+0.83}_{-0.65}$  &  \\
         & & & & & & &
         $0.44^{+0.72}_{-0.44}$  & $3.43^{+1.07}_{-0.28}$   &
         $3.28^{+0.52}_{-0.61}$  & 1.20 (3.00) \\
WR47  &      0.62     & 244/295  & 7.92   &
         $0.53^{+0.08}_{-0.08}$  & $0.59^{+0.03}_{-0.03}$  &
         $12.3^{+1.3}_{-1.2}$    &
         $0.00^{+0.21}_{-0.00}$  & $4.02^{+0.34}_{-0.33}$  &
         $8.05^{+0.41}_{-0.38}$  & 2.27 (8.02) \\
      &      0.89     &  &  &
         $0.80^{+0.11}_{-0.10}$  &  &  &
         $4.33^{+0.78}_{-0.66}$  &  &  & 1.70 (8.02)\\
WR139 &      0.47     & 362/335  & 5.86  &
         $0.23^{+0.16}_{-0.16}$  & $0.58^{+0.04}_{-0.03}$  &
         $3.66^{+0.60}_{-0.35}$  &
         $6.11^{+0.63}_{-0.58}$  & $3.08^{+0.13}_{-0.13}$  &
         $17.8^{+0.71}_{-0.62}$  & 9.40 (27.3) \\
      &      0.55     &  &  &
         $0.43^{+0.12}_{-0.15}$  &  &  &
         $2.07^{+0.29}_{-0.26}$  &  &  & 11.8( 27.3) \\
      &      0.81     &  &  &
         $0.10^{+0.09}_{-0.10}$  &  &  &
         $1.00^{+0.19}_{-0.18}$  &  &  & 13.4 (27.3) \\
WR141 &      0.02     & 35/39 & 8.20  &
         $4.40^{+0.59}_{-0.53}$  & $3.40^{+0.43}_{-0.30}$  &
         $6.82^{+0.75}_{-0.68}$   &
           & ... &  &  8.49 (19.1) \\
WR145 &      0.33     & 180/278 & $35.4^{+3.5}_{-2.2}$ &
         0.0  & $0.99^{+0.14}_{-0.12}$  & $16.5^{+3.5}_{-4.5}$  &
         0.0  & $4.75^{+1.46}_{-1.00}$  & $4.90^{+1.27}_{-1.02}$  &
         6.32 (39.7)  \\
      &      0.00     &  &  $47.5^{+8.9}_{-6.9}$ &
         0.0  &  &  &  0.0  &  &  & 4.80 (39.7) \\
      &      0.09     &  &  $57.5^{+8.4}_{-6.9}$ & 
         0.0  &  &  &  0.0  &  &  & 4.59 (39.7) \\
      &      0.18     &  &  $47.5^{+7.5}_{-5.9}$ &
         0.0  &  &  &  0.0  &  &  &  5.10 (39.7) \\
WR148 &      0.99     & 54/101 & 5.16   &
          $1.98^{+1.20}_{-0.73}$ & $0.40^{+0.18}_{-0.11}$  &
          $44.5^{+123.}_{-26.5}$   &
          $1.69^{+1.83}_{-1.69}$ & $1.03^{+1.11}_{-0.33}$  &
          $9.73^{+12.1}_{-7.64}$   & 0.255 (4.14) \\
      &      0.44     &  &  &
          $1.03^{+0.49}_{-0.50}$  &  &  &
          $10.1^{+52.4}_{-6.43}$  &  &  & 0.249 (4.14) \\
WR79  &      0.72     & 40/49 & 3.08  &
          $0.09^{+0.01}_{-0.01}$ & $1.27^{+0.48}_{-0.20}$  &
          $0.09^{+0.02}_{-0.01}$  &
          $3.92^{+1.26}_{-0.64}$ & $2.38^{+0.98}_{-0.78}$  &
          $2.92^{+4.69}_{-1.35}$  & 1.80 (33.6) \\
      &      0.23     &  &  &
          $0.09^{+0.02}_{-0.02}$ &  &  &
          $4.21^{+1.40}_{-0.78}$ &  &  &  1.70 (33.6) \\
\hline
\end{tabular}

\vspace{0.5cm}
Note --  Fit results from the two-temperature optically-thin plasma 
model with individual wind absorption for each component. 
A three-temperature plasma model was used only in the case of 
WR46: parameters of the third component are given in the second line
for this object.  The uncertainties are $1\sigma$ errors from the
fit.
 
$^{(a)}$The binary phase for each observation from Table~\ref{tab:data}.
 
$^{(b)}$The interstellar absorption column density in units 
of $10^{21}$~cm$^{-2}$.
 
$^{(c)}$The helium-dominated wind absorption column density 
in units of $10^{21}$~cm$^{-2}$.
 
$^{(d)}$The plasma temperature is in keV.
 
$^{(e)}$The emission measure ($\mbox{EM} = \int n_e n_{He}
dV $) in units of $10^{54}$~cm$^{-3}$ at the distance for each
object from Table~\ref{tab:binaries}.
 
$^{(f)}$The observed flux (0.5 - 10 keV) followed in
parentheses by the unabsorbed value. The units are 
$10^{-13}$~erg cm$^{-2}$ s$^{-1}$.

\end{table*}

\section{Discussion}
\label{sec:discussion}

Although our sample of close WR$+$O binaries is limited, the adopted
uniform way of analysing the X-ray emission from the studied objects
provides useful pieces of information to check the global consistency 
of the colliding stellar wind picture in such objects.
We note that our spectral analysis adopted simplified models: those 
of discrete-temperature optically-thin plasma (\S~\ref{sec:global_fits}). 
This means that the X-ray emitting plasma is in collisional 
ionization equilibrium (CIE), that is the non-equilibrium ionization 
(NEI) effects do not affect its emission. Also, the electron and ion
temperatures are equal. We can use the corresponding
characteristic plasma timescales to check validity of these
assumptions in the case of close WR$+$O binaries.

For this purpose, two dimensionless parameters have been introduced 
$\Gamma_{eq} = \tau/\tau_{eq}$ \citep{zhsk_00} and
$\Gamma_{NEI} = \tau/\tau_{NEI}$ \citep{zhek_07}, where $\tau$ is the
timescale of the gasdynamics, $\tau_{eq}$ is the
electron-ion temperature equalization time and $\tau_{NEI}$ is the 
representative NEI timescale. We recall that the electron and ion
temperatures are different if $\Gamma_{eq} \leq 1 $
but their difference can be neglected if $\Gamma_{eq} \gg 1 $. 
Similarly, the NEI effects must be taken into account if 
$\Gamma_{NEI} \leq 1 $ but can be neglected if $\Gamma_{NEI} \gg 1 $.

For a total mass of the WR$+$O systems of 30-40~M$_{\odot}$, we can
use the Kepler's third law 
($a = 2.928\times10^{11} P^{2/3}_d [M/M_{\odot}]^{1/3}$~
cm; $P_d$ is the orbital period in days (Table~\ref{tab:binaries}); 
$M$ is the total mass) 
to estimate the typical lengthscale for the gasdynamic problem
(e.g., the binary separation). Then the stellar wind parameters from 
Table~\ref{tab:binaries} ($\dot{M}, V_{\infty}$) along with
eq.(1) in \citet{zhsk_00} and eq.(1) in \citet{zhek_07} show
that, first, the plasma in the CSW shock in the close WR$+$O 
binaries considered here has equal electron and ion temperatures
($\Gamma_{eq} \gg 1 $). Second, the NEI effects are not important in 
these objects ($\Gamma_{NEI} \gg 1 $).
All this makes us confident that the adopted spectral models in this
study were adequate to the physical conditions of the X-ray emitting
plasma in close WR$+$O binaries.

On the other hand, when fitting the total (`integrated') X-ray
emission from temperature stratified plasmas (e.g., present in
the interaction region in CSW binaries) with discrete-temperature
models, we can expect that the derived plasma temperatures will not be
higher than the maximum one in the X-ray emitting region. We note that
for a strong shock (with adiabatic index $\gamma = 5/3$) in
helium-dominated plasma (mean molecular weight per particle 
$\mu = 4/3$) the postshock temperature
is kT $= 2.608 (V_{sh}/1000 \mbox{\kms})^2$~keV, where
$V_{sh}$ is the shock velocity, and the maximum shock velocity is equal 
to the stellar wind velocity ($V_{sh} = V_{\infty}$) in the framework of 
CSW model. We thus see from Tables~\ref{tab:binaries}
and~\ref{tab:fits} that the deduced plasma temperatures in our sample
of close WR$+$O binaries are qualitatively consistent with the adopted 
physical picture. Namely, all of them are lower than the maximum 
possible plasma temperature in the interaction region of these CSW 
objects.

Also, in the CSW picture the hotter plasma is located near the line
connecting the two stellar components in the binary system, that is in
the denser part of the stellar winds. Thus, we can expect that the 
higher-temperature component of the discrete-temperature models will 
suffer higher `wind' absorption. As seen from Table~\ref{tab:fits},
this is in general found in the spectral fits to the X-ray emission
from the close WR$+$O binaries in our sample (although there are some
exceptions  but likely due to the poor photon statistics). 

It then seems conclusive that the X-ray emission from  the 
close WR$+$O binaries considered in this study is qualitatively
consistent with the physical picture where X-rays arise in hot
temperature-stratified plasmas behind colliding stellar wind shocks.
But, an interesting issue is worth discussing as well, namely, the
total X-ray energetics (luminosity) of CSWs in close binaries which 
is directly related to whether the CSW shocks are radiative, 
adiabatic etc.

\begin{figure*}
 \centering\includegraphics[width=2.8in,height=2.in]{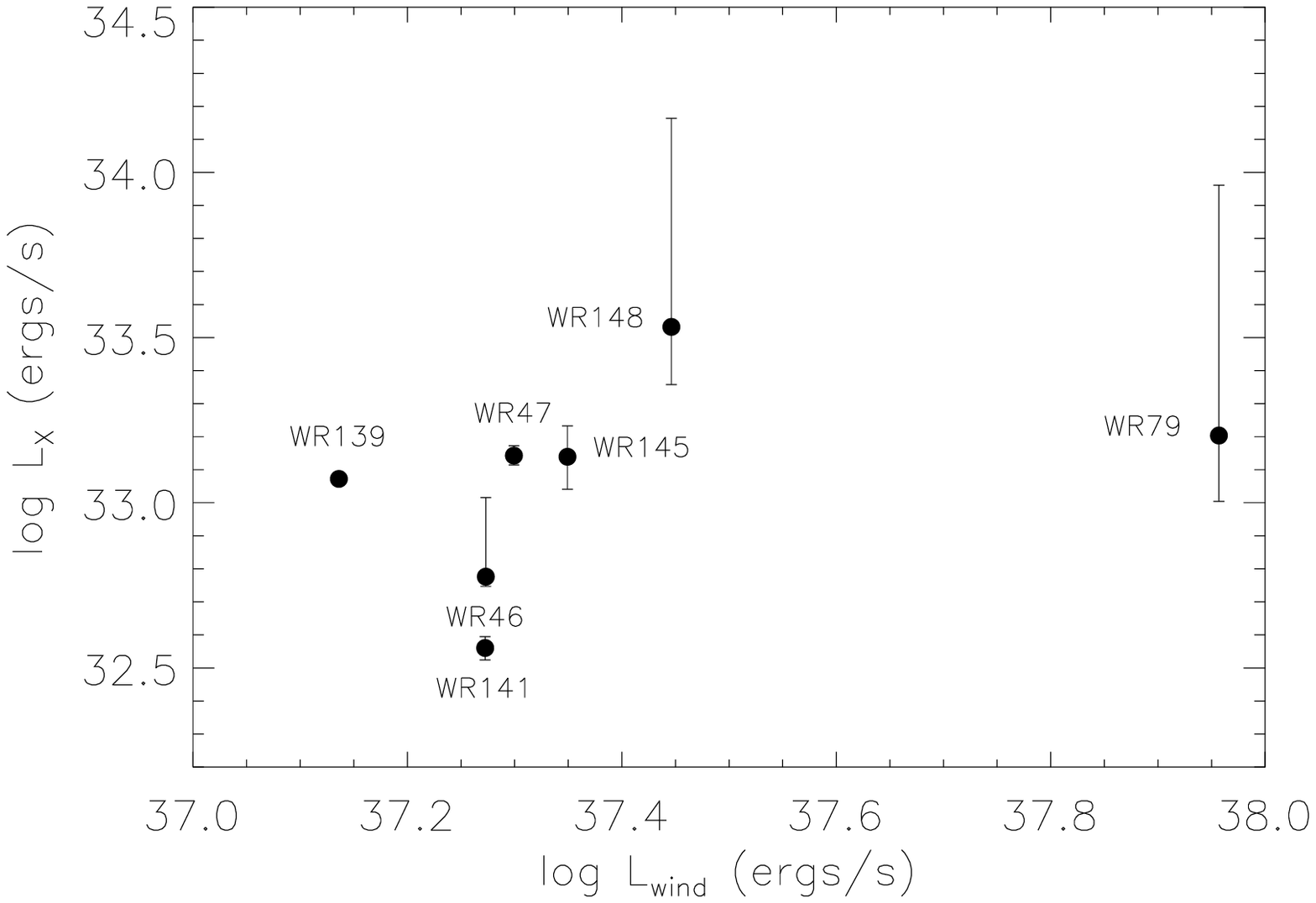}
 \centering\includegraphics[width=2.8in,height=2.in]{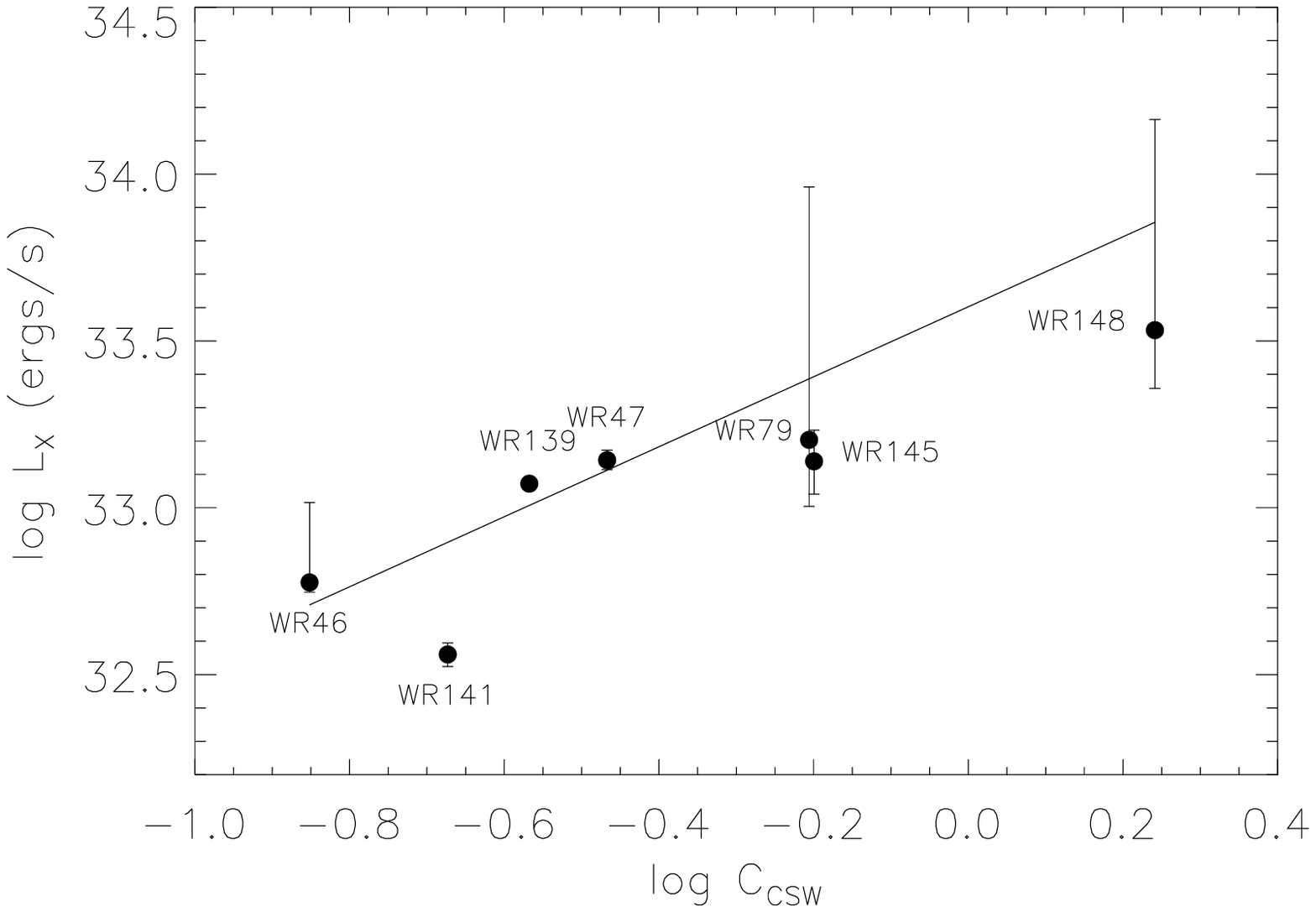}
\caption{
{\it Left} panel: X-ray (L$_X$) vs. wind luminosity 
(L$_{wind} = \frac{1}{2} \dot{M} V_{\infty}^2 $) for the sample of
close CSW binaries.
{\it Right} panel: X-ray luminosity vs. the CSW parameter,
$C_{CSW} = (\dot{M}/10^{-5}\mbox{\msolr})^2 
(V_{\infty}/ 1000 \mbox{\kms})^{-3} P_d^{-2/3}$.
The slope of the solid line is $1.05\pm0.09$.
The X-ray luminosity is calculated for the corresponding distance
to each object (Table~\ref{tab:binaries}). The error bars on
$L_X$ are internal only and do not take into account those on distance
to the objects.
}
\label{fig:csw}
\end{figure*}

We can use the dimensionless parameter $\chi = \tau_{cool}/\tau$
introduced by \citet{st_92}  to estimate the
importance of radiative losses behind the CSW shocks in the WR$+$O
binaries of our sample. We recall that $\tau_{cool}$ and $\tau$ are
the characteristic cooling time of the shocked plasma and the timescale 
of the gasdynamics, respectively. Then from the data in
Table~\ref{tab:binaries} and eq.(8) in \citet{st_92}, we see
that the radiative cooling should have an important effect on the 
physics of CSWs in the close WR$+$O binaries studied here ($\chi < 1$).
This means that only numerical hydrodynamic modelling of the
interaction region must be used to confront theory and observations.
However, we can perform a qualitative check on the X-ray energetics,
as mentioned above, by considering two extremes:
(a) highly radiative and (b) adiabatic CSW shocks.

In the case of radiative shocks, we have an upper limit on the available 
energy (luminosity) that can be converted into X-ray emission: no more 
energy is emitted than the energy flux crossing the shock front per unit 
area. For the maximum X-ray luminosity, we have L$_X = \frac{1}{2}\int
\rho V_{wind,\perp}^3 dS$ 
($\rho$ is the density of the wind in front of the shock; 
$V_{wind,\perp}$ is the wind velocity component perpendicular to the 
shock front; $S$ is the shock surface). In the case of CSWs, the 
integration is over the entire CSW `cone' and although its shape
varies from one object to another we can expect that a relation $L_X \propto
L_{wind}$ might hold in general for highly radiative CSWs
($L_{wind}=\frac{1}{2}\dot{M} V_{\infty}^2$ is the stellar wind
luminosity).

On the other hand, in the case of adiabatic CSWs there exists a scaling 
law for the CSW X-ray luminosity with the mass-loss rate ($\dot{M}$), 
wind velocity ($V_{\infty}$) and binary separation ($a$): 
$L_X \propto \dot{M}^2 V_{\infty}^{-3} a^{-1}$ (\citealt{luo_90}; 
\citealt{mzh_93}). So, using the third Kepler's law we can
write: $L_X \propto C_{CSW}$ by introducing 
$C_{CSW} = (\dot{M}/10^{-5}\mbox{\msolr})^2 
(V_{\infty}/ 1000 \mbox{\kms})^{-3} P_d^{-2/3}$ as a
'colliding stellar wind' parameter.

Figure~\ref{fig:csw} presents two plots: the X-ray vs. the wind luminosity 
and the X-ray luminosity vs. the CSW parameter for the objects in our
WR$+$O binaries sample (the unabsorbed fluxes from 
Table~\ref{tab:fits} are used for calculating the X-ray luminosity). 
We see that there is no correlation for the
former while the X-ray luminosity clearly correlates with the CSW
parameter. Interestingly, the derived proportionality between 
$\log L_X$ and $\log C_{CSW}$ is $1.05\pm0.09$ (the error is $1\sigma$ 
error from the fit), thus, $\L_X \propto C_{CSW}^{1.05\pm0.09}$. 
All this is quite surprising since it means that the CSW
shocks in close WR$+$O binaries are more likely adiabatic rather than 
being strongly radiative. We note again that our sample of studied
objects is limited but nevertheless the results in Fig.~\ref{fig:csw} 
are  very interesting and we will discuss them in some detail.

\begin{figure*}
 \centering\includegraphics[width=2.8in,height=2.in]{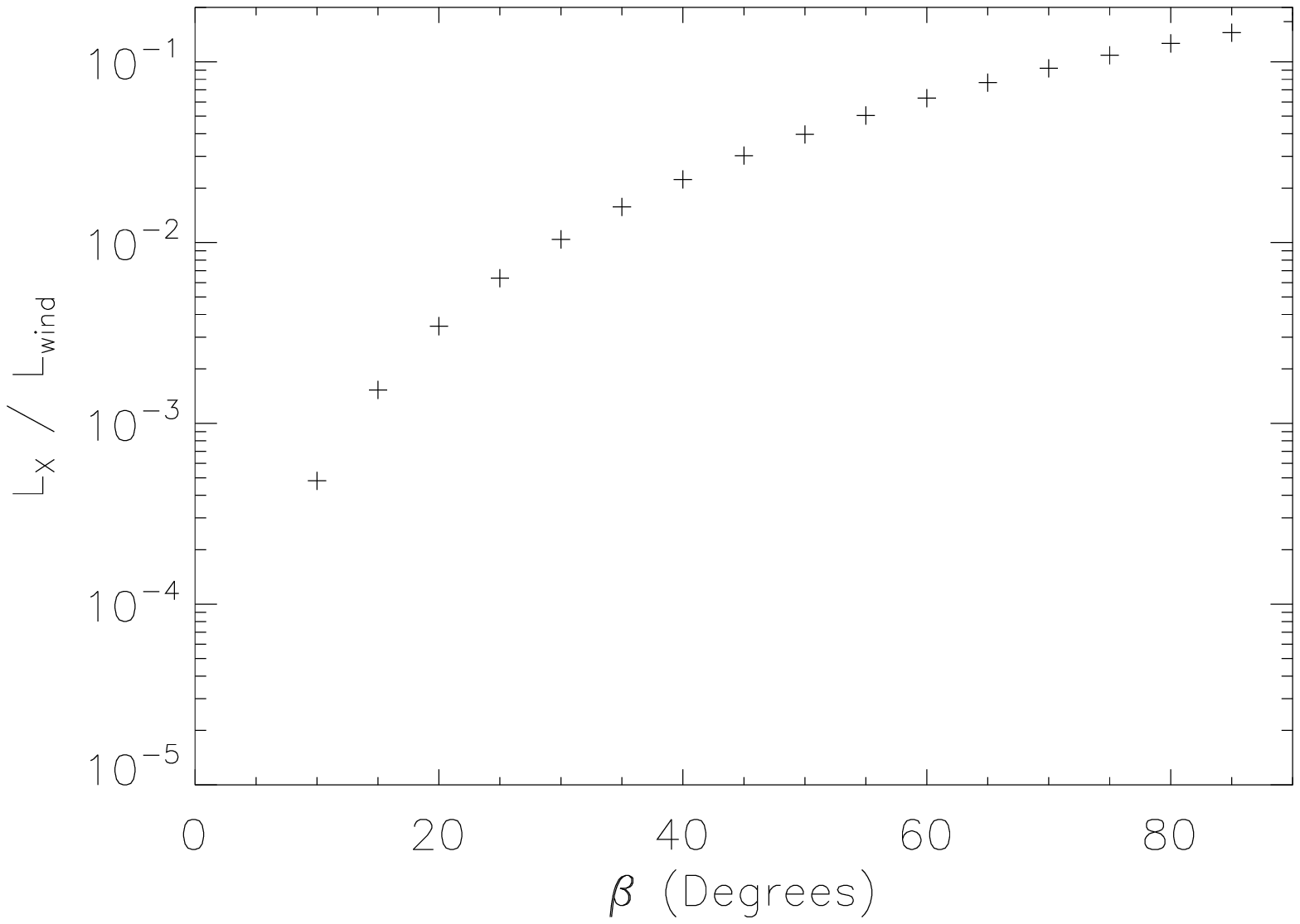}
 \centering\includegraphics[width=2.8in,height=2.in]{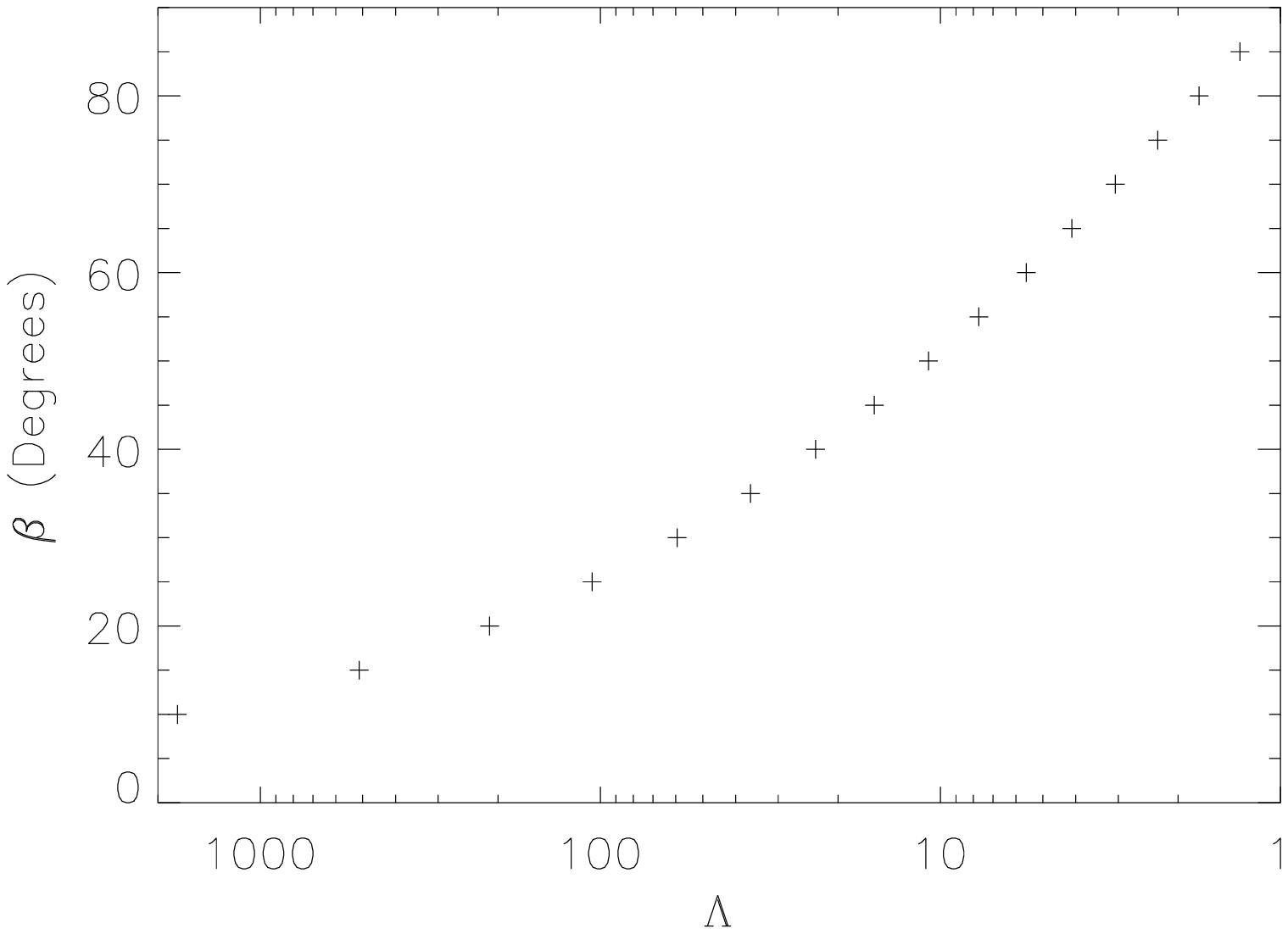}
\caption{
{\it Left} panel: the ratio of the maximum possible X-ray luminosity 
(L$_X$) to the wind luminosity 
(L$_{wind} = \frac{1}{2} \dot{M} V_{\infty}^2 $) for different
values of the half-opening angle ($\beta$) of the CSW `cone' in the case
of highly radiative CSW shocks in binary system.
{\it Right} panel: dependence of the half-opening angle ($\beta$) on the
ratio of the ram pressure of the stellar winds in the binary system
($\Lambda = \dot{M}_{WR} V_{WR} / \dot{M}_{O} V_{O}$, where $\dot{M}_{WR}, 
V_{WR}, \dot{M}_{O}, V_{O}$ are the wind parameters, mass loss rate and 
wind velocity, of the WR and O star, respectively).
}
\label{fig:lx_lwind}
\end{figure*}

It is worth noting another result from Fig.~\ref{fig:csw} that favours
the case of adiabatic CSWs in the close WR$+$O binaries in our sample. 
Namely, the low efficiency of converting the wind luminosity into X-ray 
emission: $L_X/L_{wind} \sim 10^{-4}-10^{-5}$. Of course, this ratio
cannot be very close to unity because the CSW shock cone occupies only
part of the `sky' of the WR star (the shocked WR wind dominates the
X-ray emission from the interaction region). But, it seems unlikely to
explain such a low value of $L_X/L_{wind}$ only by geometry effects and
some other explanation is needed. Moreover, by its very definition the
term `radiative shocks' means that high percentage of the energy
influx is radiated away (mostly in X-rays for fast shocks), 
while the opposite is valid for `adiabatic shocks'. 
Figure~\ref{fig:lx_lwind} shows the result from our numerical 
integration for the maximum possible X-ray luminosity, 
L$_X = \frac{1}{2}\int \rho V_{wind,\perp}^3 dS$, in
the case of highly radiative CSW shocks. In such a case, the
interaction region `collapses' and its shape coincides with that of 
the contact discontinuity between the shocked WR and O star winds. 
We adopted the \citet{canto_96} solution for the shape of the CSW
`cone'. We see that the theoretical $L_X/L_{wind}$ values are
considerably higher than those observed.
Thus, the relatively low values for
the X-ray luminosity, with respect to the wind luminosity of 
the more massive star in the binary, do support the conclusion that 
the CSWs in our sample of close WR$+$O binaries are {\it adiabatic}.

But, adiabatic CSW shocks are possible in this case {\it only} if the
mass-loss rate of the stellar winds in the binaries is considerably
smaller, e.g. at least by an order of magnitude, than the values
listed in Table~\ref{tab:binaries} (we note that in such a case the
actual values of the CSW parameter will be smaller and the plot in the
right panel of Fig.~\ref{fig:csw} will be shifted to the left in
the x-axis; a detailed analysis in that case is needed: see discussion 
below). In other words, our analysis of
the X-ray emission from close  WR$+$O binaries finds an indication
of considerable clumping in the winds of these massive stars. 
It then seems plausible to propose that the stellar wind of a massive
WR (O) star is likely a two-component flow: a rarefied continuous
component with low mass-loss rate and a `discrete' component
consisting of numerous dense clumps. The first component forms the
CSW interaction region, thus, it plays role for
the X-ray emission from close WR$+$O binaries while the second one
dominates their optical/UV spectra. We note that in the last two
decades the physical picture of clumpy stellar winds in the hot 
massive stars gathered observational evidence and has
become a standard ingredient in the analysis of the optical, UV and
infrared spectra of these objects (see \citealt{puls_08} for a recent
review on the matter). Here from the analysis of X-ray spectra of
close WR$+$O binaries, we find additional evidence that the stellar wind 
in massive stars is clumpy (highly inhomogeneous) on a 
lengthscale at least of the order of the binary separation
(approximately the size of the CSW region). It is worth noting though
that we need more detailed knowledge, both observational and theoretical, 
in order to build a coherent physical picture of clumpy stellar winds in 
massive stars. For example, what is the origin of the clumps; what is the 
distribution of clumps by their size; what is the clump evolution with 
the distance from the star; do the homogeneous  component of the wind and 
the clumps share the same bulk velocity; thus, what is the efficiency
of the radiative force that drives both of them;
do both components of the stellar wind reach terminal wind
velocity in front of the interaction region; does radiative braking
play an important role for this complex wind structure?

An interesting issue in this respect is about the fate of the dense
clumps when interacting with the CSW region of the rarefied
homogeneous components of the stellar winds in close WR$+$O binaries.
\citet{cherep_90} proposed a qualitative picture where the dense
clumps pass freely through the interaction region and part of them are
decelerated in the O-star photosphere where their X-rays are 
transformed into optical emission. On the other hand,
\citet{pittard_07}
carried hydrodynamical simulations of a CSW binary with clumpy winds
that showed efficient destruction of clumps while interacting with the
CSW shocks. As a result, the average density in the interaction region
increases and is similar to the case of smooth winds with the higher 
mass-loss rate. We have to keep in mind that these simulations were
suitable for wide CSW binaries and specifically for the stellar wind
and binary parameters of WR140 (see \citealt{pittard_07} for details) 
and the reality might be quite different in close CSW binaries. Namely, 
the clumps are likely smaller closer to the base of the stellar wind, 
they may have much larger density contrast with respect to the smooth
component etc. Also, if the clumps are destroyed in the CSW region in
close WR$+$O binaries, then these objects would have been much more
luminous in X-rays than deduced from observations. Thus, it seems
plausible to assume that the dense clumps of the stellar winds
`survive' while crossing the CSW region in close WR$+$O binaries.
It is difficult to describe what could be the possible observational
evidences from the interaction of these dense clumps with the CSW
region since they depend on such `hard-to-guess' details as those
mentioned at the end of the previous paragraph. 

We thus believe that
future X-ray observations with much better photon statistics might be
very helpful in this respect. Results from such observations must be
considered in conjunction with those from the 
optical/UV spectral domain to build a self-consistent physical picture 
of the stellar winds in massive stars. Such a global analysis may
reveal that the smooth component has very little or no contribution to
the optical/UV emission of a massive star and X-ray observations could
be the only tool to reveal its presence and physical properties. 
It should also take into account that massive O stars are X-ray sources 
themselves (for a recent review see \citealt{gudel_naze_09}). We just 
note that their emission is rather `soft' with plasma temperatures 
below 1 keV (e.g., \citealt{woj_05}; \citealt{zhp_07}; see also 
Sections 4.1.3 and 4.3 in the review paper of \citealt{gudel_naze_09}), 
opposite to what is found in close WR$+$O binaries (e.g., 
Table~\ref{tab:fits}). On the other hand, the issue of the intrinsic
contribution of the WR star to the total X-ray emission from the binary 
might still bear a lot of uncertainties.

Pointed observations with modern X-ray observatories ({\it Chandra; 
XMM-Newton})  of a few presumably single WC stars resulted only in 
non-detections (\citealt{osk_03}; \citealt{sk_06}). Could it be that 
all single WC stars are X-ray quite? In such a case, there will be no 
contribution from a WC star to the total X-ray emission from a WR$+$O 
binary system.

In contrast to this, the \citet{sk_10} analysis of the X-ray 
spectra  of a small sample of presumably single WN stars showed that 
these objects emit X-rays that arise from an admixture of cool 
(kT $< 1$~keV) and hot (kT $> 2$~keV) plasma. Presence of hot plasma
(kT $> 2$~keV) makes them similar to the CSW binaries studied here and
distinct from the single O stars. 
{\it
So, could it be that these WN stars are not single but binaries
instead?
}
However, there are some differences in the X-ray characteristics 
between the single WN stars and the close WR$+$O
binaries in our sample. Namely, the latter are X-ray more luminous
with a minimum value of $\log L_X \approx 32.5$ \ergs
(Fig.~\ref{fig:csw}) while most of the X-ray detected presumably single
WN stars have luminosities below this figure (see Fig.~10 in
\citealt{sk_10}). Also, there is indication that the 
$L_X \propto L_{wind}$ 
relation holds for single WNs while this is not the case for close
WR$+$O binaries (compare Fig.~10 in \citealt{sk_10} with the left
panel in Fig.~\ref{fig:csw} here).
Since the X-ray production mechanism in single WN stars has not been
identified yet, one possibility is that those of them detected in
X-rays are in fact WR binaries with a normal (non-degenerate) companion
(see discussion in \S~4.4 in \citealt{sk_10}) which may also
explain the $L_X \propto L_{wind}$ trend for them (see
\S~7.2 in \citealt{zhp_10b}).
And we emphasize again that more X-ray data with good quality are
needed and the global analysis as that mentioned above
will help us better understand the physical picture in the stellar
winds of hot massive stars.

Finally, the object with the shortest period in our sample, WR46,
deserves a few more comments. This object has a very complex
variability pattern in the optical, UV and may be X-rays (e.g.,
\citealt{mar_00}; \citealt{goss_11a}; \citealt{hen_br_11};
but see also Appendix~\ref{app} here). {\citet{goss_11a} proposed 
that its hard X-ray emission may arise in CSW shocks. 
\citet{hen_br_11} argued for non-radial pulsations as the
most likely scenario to explain its characteristics and they 
even drew analogy between WR46 and $\zeta$ Pup (a massive presumably
single O star) in this respect.
Based on the results presented here (e.g., high plasma temperatures)
and since the X-ray variability pattern is not well established (see 
Appendix~\ref{app}), we believe that the physical picture of CSWs in 
a short-period WR$+$O binary is a more likely explanation for the 
X-ray properties of WR46.

\section{Conclusions}
\label{sec:conclusions}

Using data from the {\it Chandra} and {\it XMM-Newton} public archives, 
we have analysed the X-ray emission from a small sample of close WR$+$O
binaries. In such objects, X-rays likely originate in colliding stellar 
wind shocks driven by the massive winds of the binary components. The 
main results and conclusions from our analysis are as follows.

(1) Global spectral fits show that two-temperature plasma is needed 
to match the X-ray emission from these objects as the hot component
(kT~$> 2$~keV) is an important ingredient of the spectral models.

(2) In general, CSW shocks in close binaries are expected to be 
radiative due to the high density of the plasma in the interaction
region and the X-ray emission is dominated by that of the shocked WR
wind. Thus, a correlation between the X-ray luminosity from these
objects and the mechanical luminosity of the WR wind should exist:
$L_X \propto L_{wind}$. Interestingly, we do not find such a
correlation for the objects studied here. 

(3) Our analysis shows that a correlation between 
the X-ray luminosity and the so called CSW parameter
(see \S~\ref{sec:discussion}) does hold. This means that 
CSWs in close WR$+$O binaries must be adiabatic and this
is possible only if the mass-loss rates of the hot stars in these
objects are at least one order of magnitude smaller than the values
currently accepted. 

(4) The most likely explanation for the X-ray 
properties of close WR$+$O binaries could be that their winds are 
two-component flows.  The more massive component (dense clumps) play 
role for the optical/UV emission from these objects. However, the smooth 
rarefied component is a key factor for their X-ray emission. We believe 
that global analysis (modelling) of optical, UV, X-ray emission from 
close WR$+$O binaries will help us build a self-consistent physical 
picture of the stellar winds and the close circumstellar environment in 
these objects.

(5) To further check the results presented here, similar analysis should
be applied to close O$+$O binaries since they are evolutionary progenitors
of the WR$+$O binaries. The work by \citet{de_be_04} is an example for the 
relevance of such a proposition. Their study of the X-ray
emission from HD159176 (an O$+$O system with a 3.367-day period)
showed that the CSW model overestimates the observed X-ray luminosity
for the standard wind parameters.

\section{Acknowledgments}
The author acknowledges financial support from Bulgarian National 
Science Fund grant DO-02-85.
This research has made use of the NASA's Astrophysics Data System, and
the SIMBAD astronomical data base, operated by CDS at Strasbourg,
France.
Also, the author thanks the referee Eric Gosset for valuable comments 
and suggestions.

\appendix
\section{Details on the X-ray data for WR46 and WR79}
\label{app}

\subsection {WR46}

\begin{figure*}
 \centering\includegraphics[width=2.in,height=1.5in]{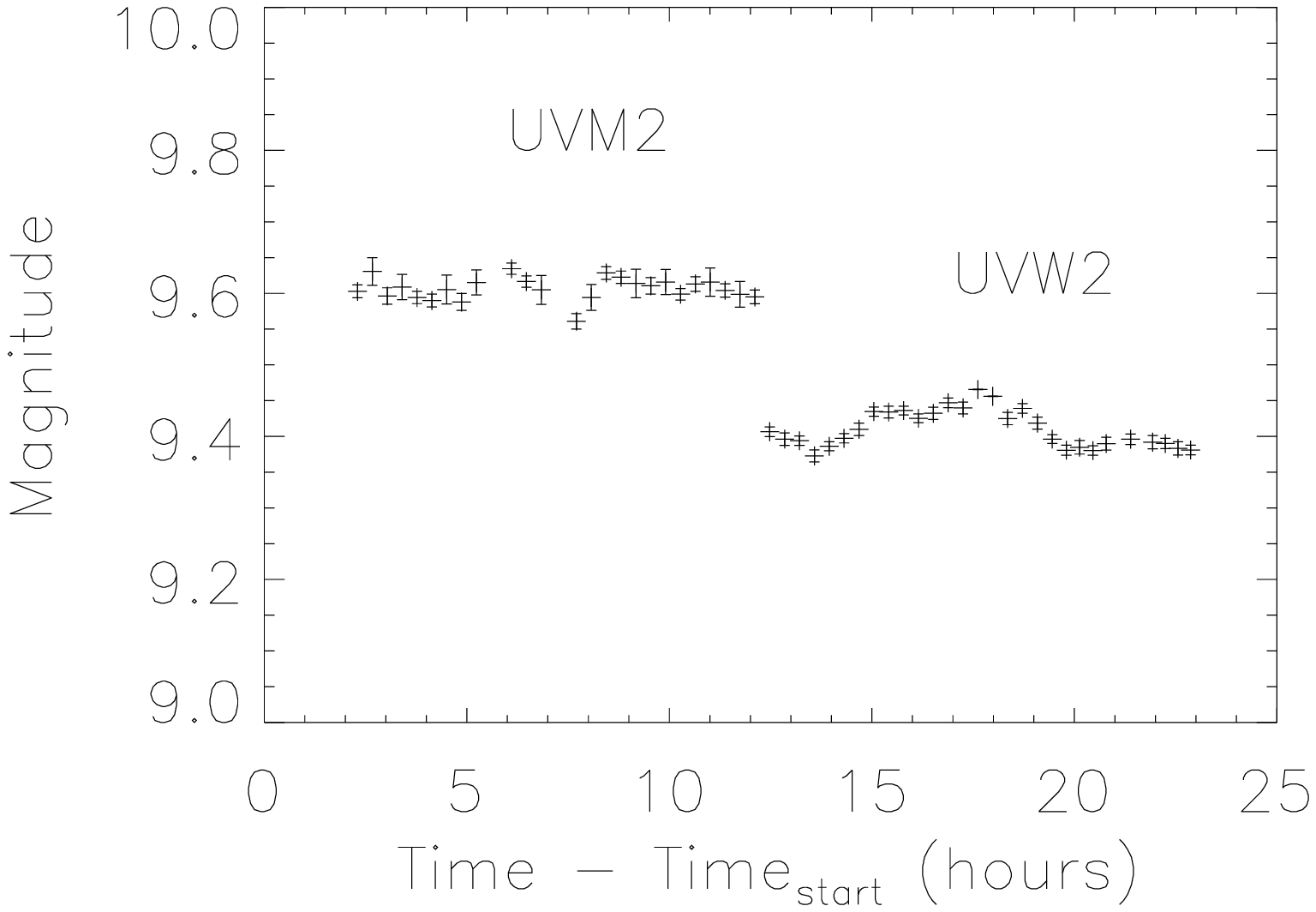}
 \centering\includegraphics[width=2.in,height=1.5in]{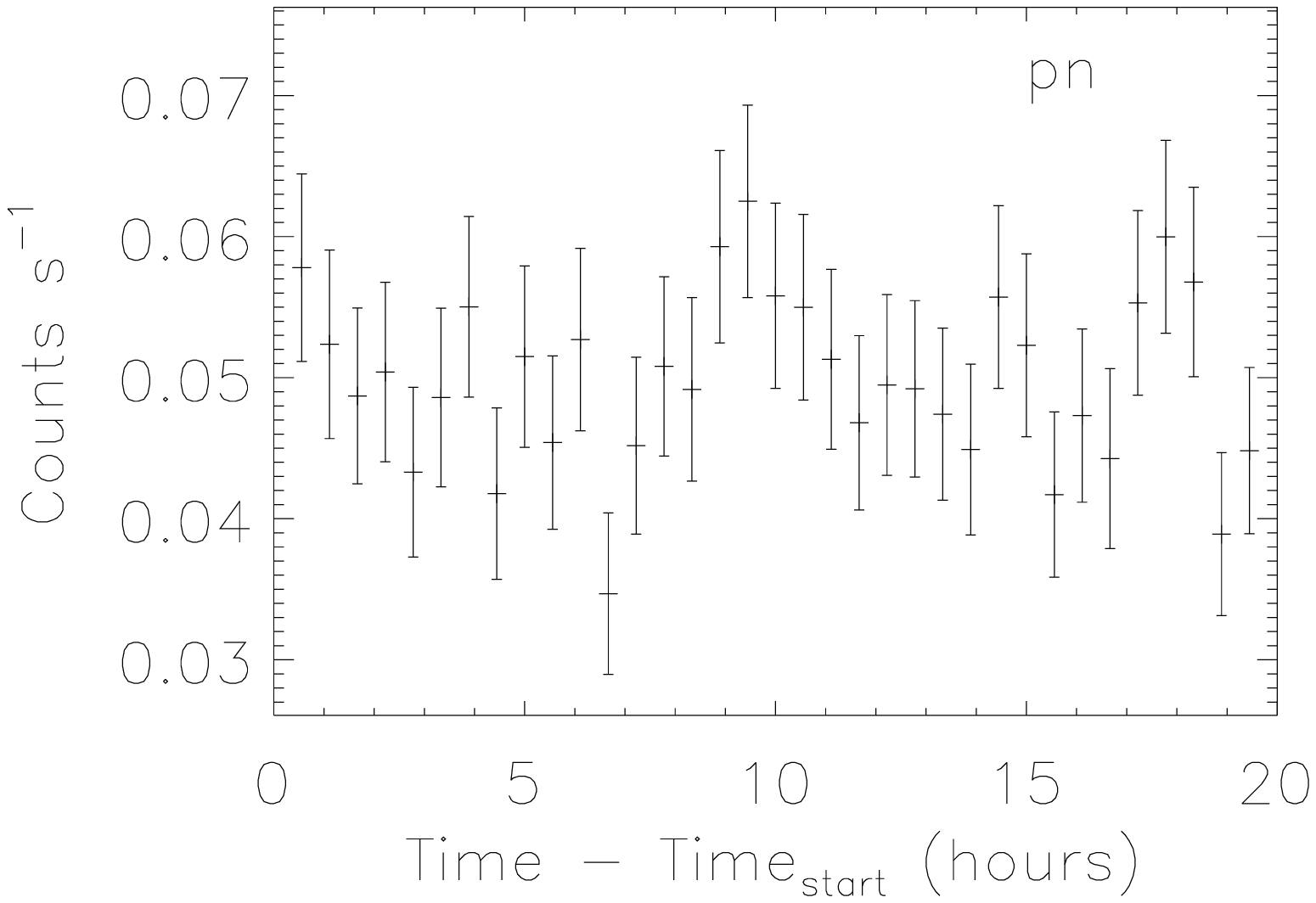}
 \centering\includegraphics[width=2.in,height=1.5in]{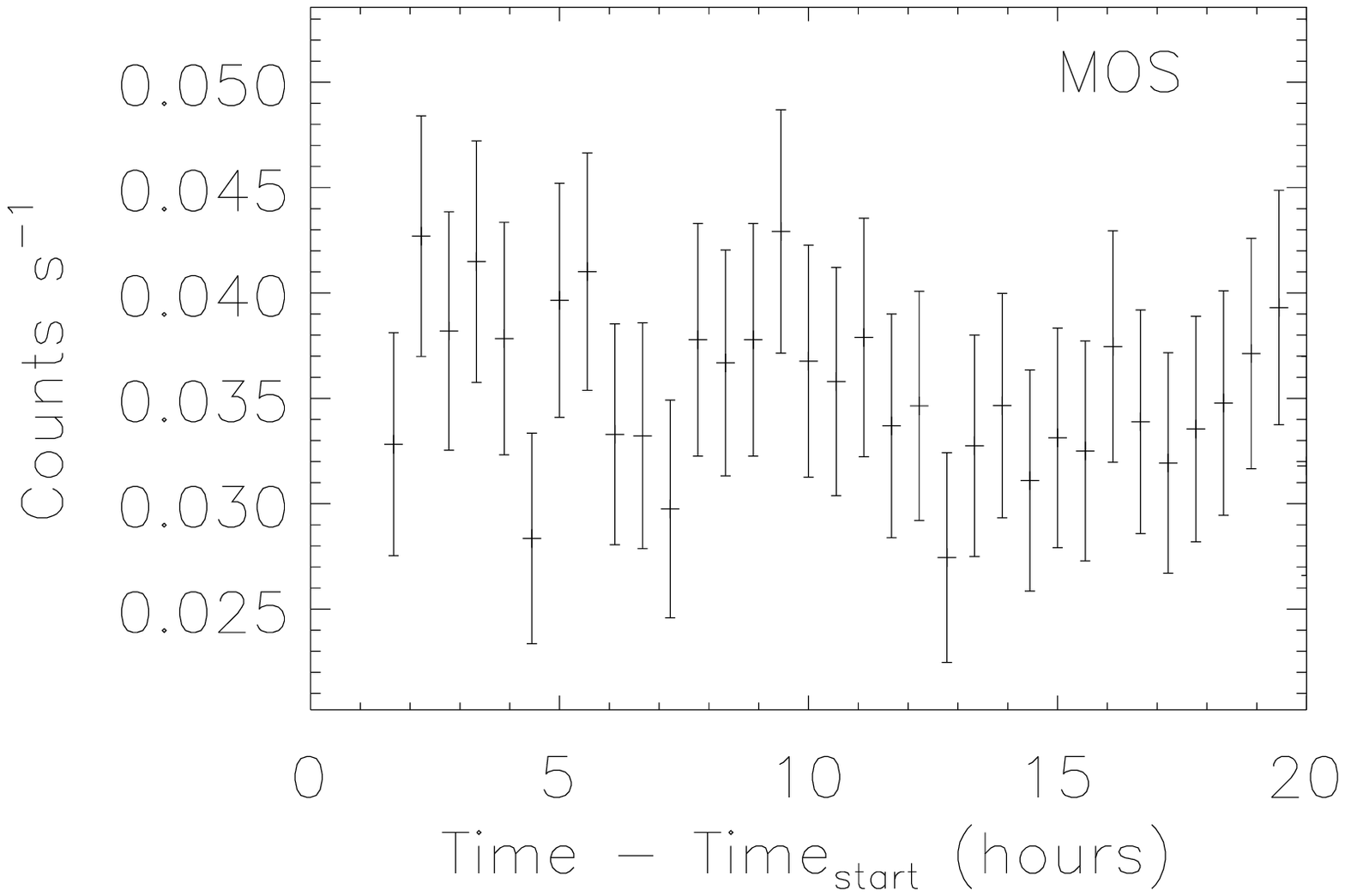}
 \centering\includegraphics[width=2.8in,height=2.0in]{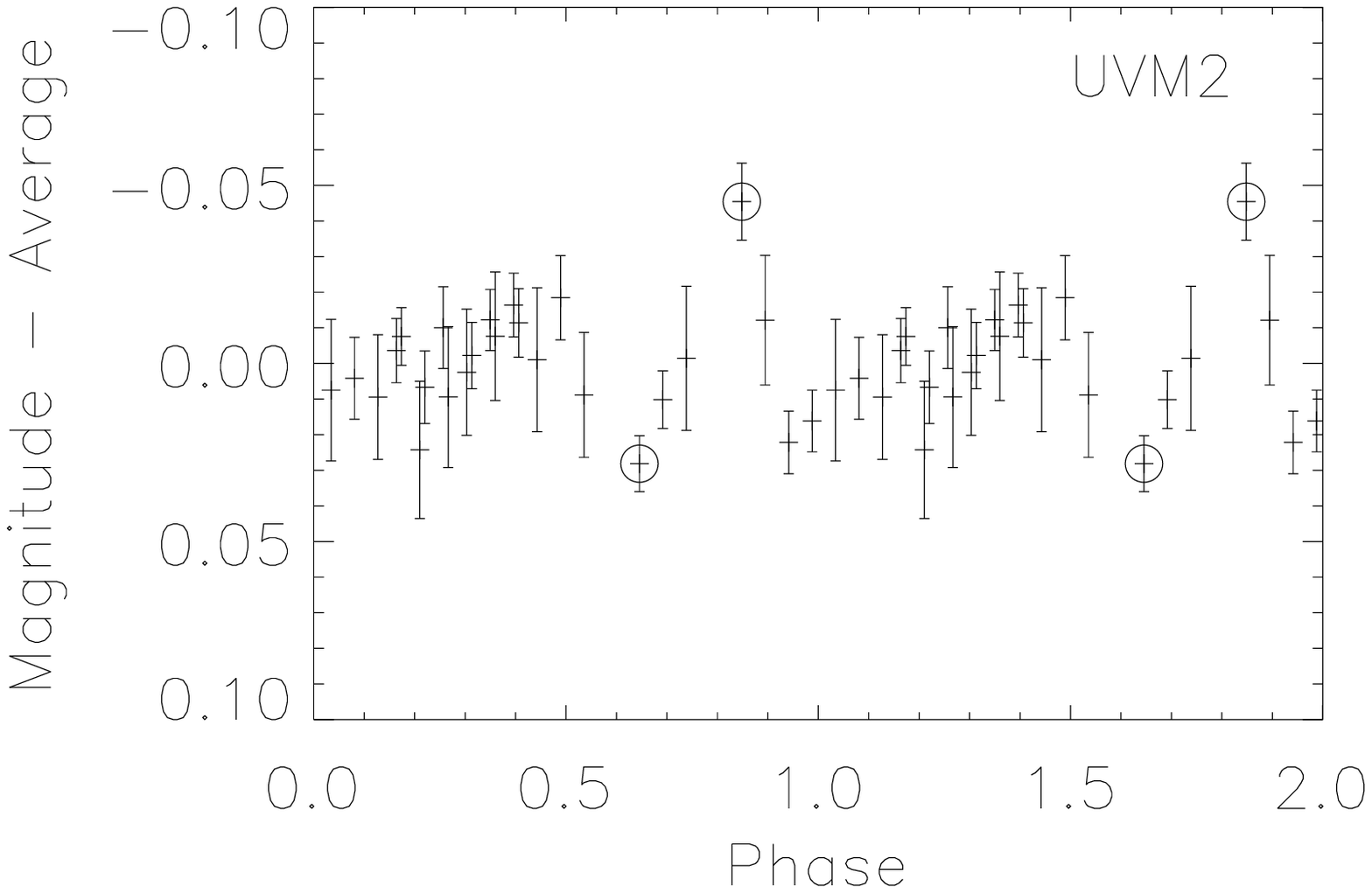}
 \centering\includegraphics[width=2.8in,height=2.0in]{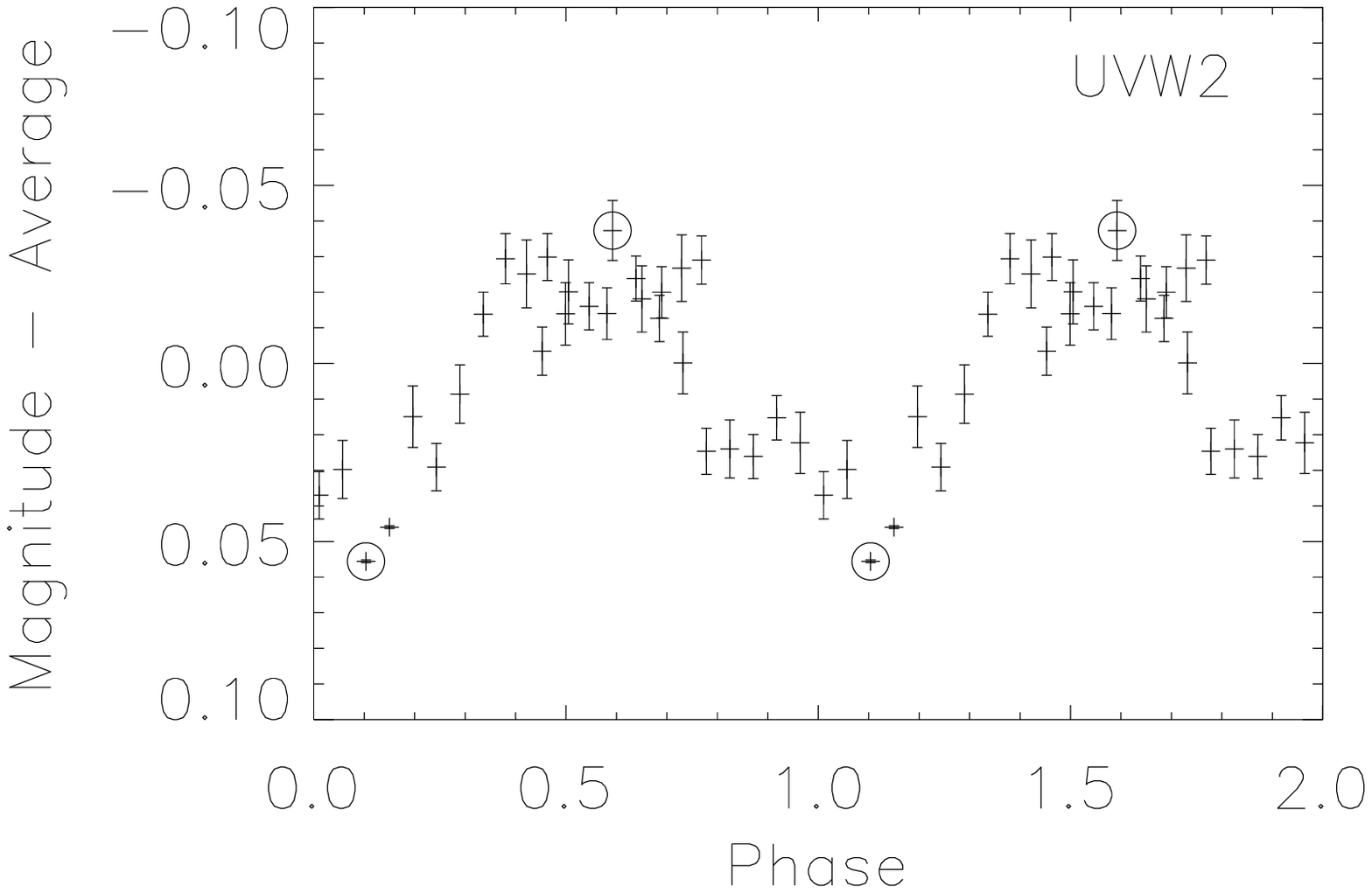}
 \centering\includegraphics[width=2.8in,height=2.0in]{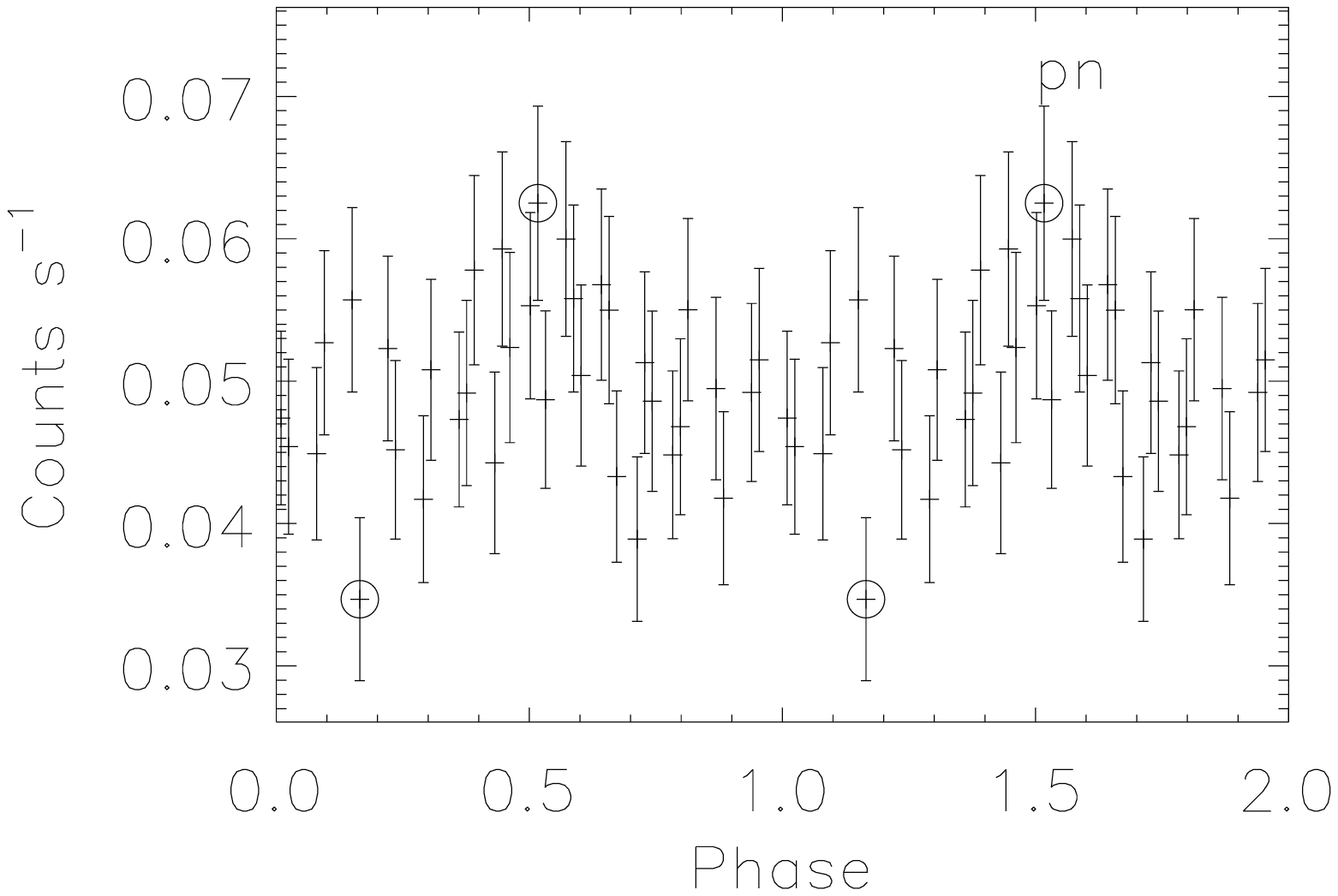}
 \centering\includegraphics[width=2.8in,height=2.0in]{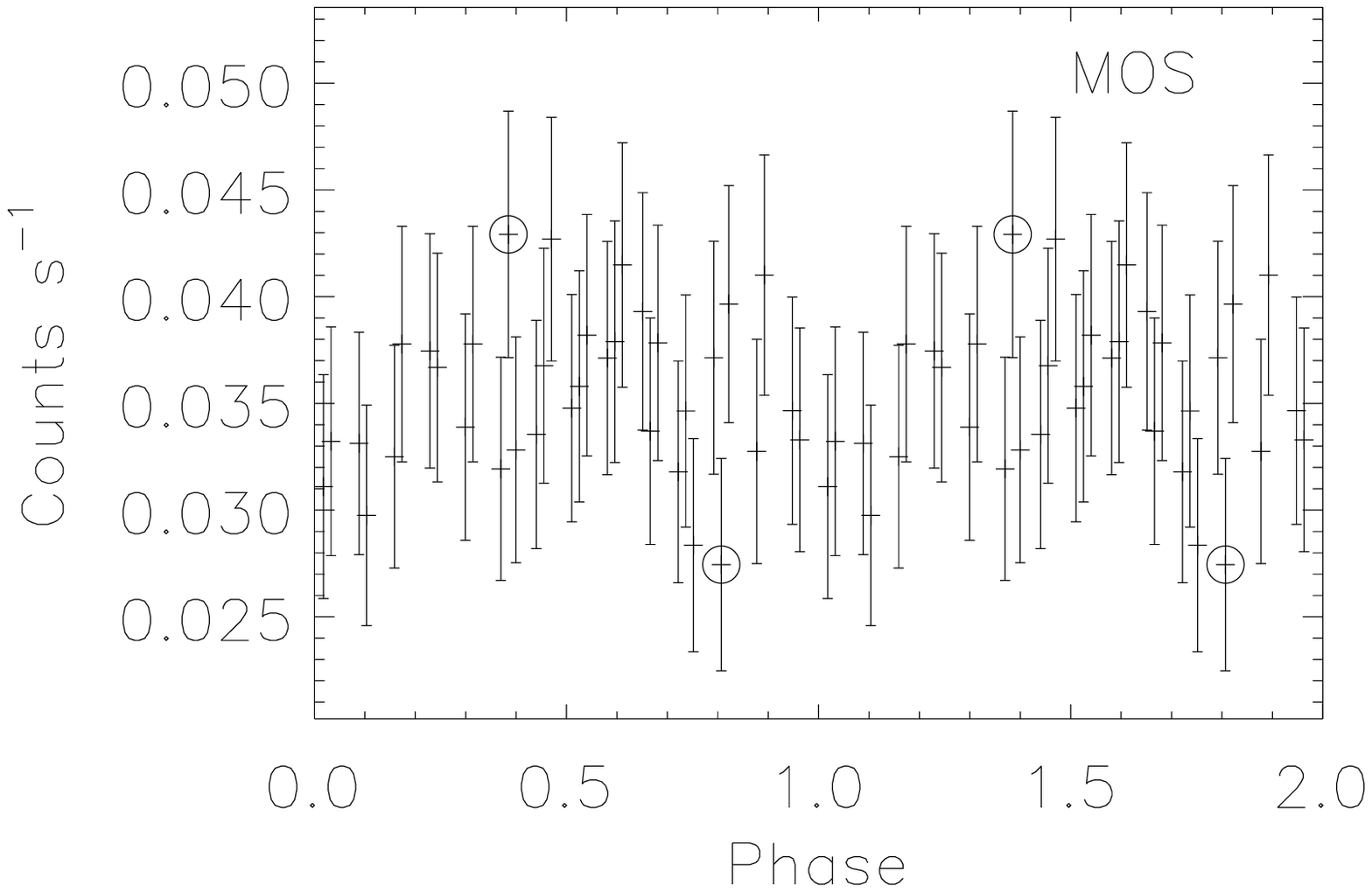}
\caption{
{\it XMM-Newton} light curves (LC) of WR46.
{\it First row:} LCs as a function of observing time in two UV
filters of the Optical Monitor (UVM2; UVW2) and in the (0.2 - 10 keV)
energy range with the EPIC detectors (pn; MOS denotes the total emission 
from the two MOS detectors). 
{\it Middle and Bottom rows:} LCs folded with the orbital period;
the values for period and zero-phase time are from \citet{mar_00}. 
Note that for graphical convenience the data points are plotted twice: 
the data for phase values (1.0-2.0) are the same as for phase values 
(0.0-1.0). Circles mark the maximum and minimum values in each LC.
The X-ray LCs are background-subtracted.
}
\label{fig:wr46}
\end{figure*}

One of the most interesting characteristics of WR46 is its variability
established in the optical/UV (e.g., \citealt{mar_00} and the
references therein). Since {\it XMM-Newton} is capable of providing
optical/UV data simultaneously with the X-ray data,
the corresponding WR46 observations allow to search
for X-ray variability that may correlate with the one found in the
optical/UV. \citet{goss_11a} and \citet{hen_br_11}
reported that such a correlation does exist (we note that
\citealt{goss_11a} suggested that the X-ray variability is
restricted only to the 0.2-0.5 keV energy range).

Adopting a recent version of the SAS software (\S~\ref{sec:data}), we 
extracted background-subtracted X-ray light curves (LC) from the 
reprocessed EPIC data. We also built UV LCs in the two filters
(UVM2, UVW2) from the reprocessed data of the Optical Monitor used in 
these observations.
For consistency with the UV data, the X-ray LCs are binned in
2000-second time intervals.
Figure~\ref{fig:wr46} presents the corresponding X-ray and UV LCs of
WR46. We see that the variability pattern in WR46 is quite complex. 
Namely, the expected 0.329-day period is evident {\it only} in the UV
filter UVW2. This is not the case for the other UV filter (UVM2)
although these data are taken one after the other (see the left panel
in the first row in Fig.~\ref{fig:wr46}). Such a behaviour is quite 
strange (or interesting) since the throughput curves for these filters 
overlap considerably (see Fig.88, \S~3.5.5.1 in the XMM-Newton Users 
Handbook\footnote{see Documents \& Manuals on http://xmm.esac.esa.int/}).
Similarly, the X-ray variability (if any) does not seem to be 
`identical' in the EPIC detectors of a different kind: pn and MOS.

\begin{figure*}
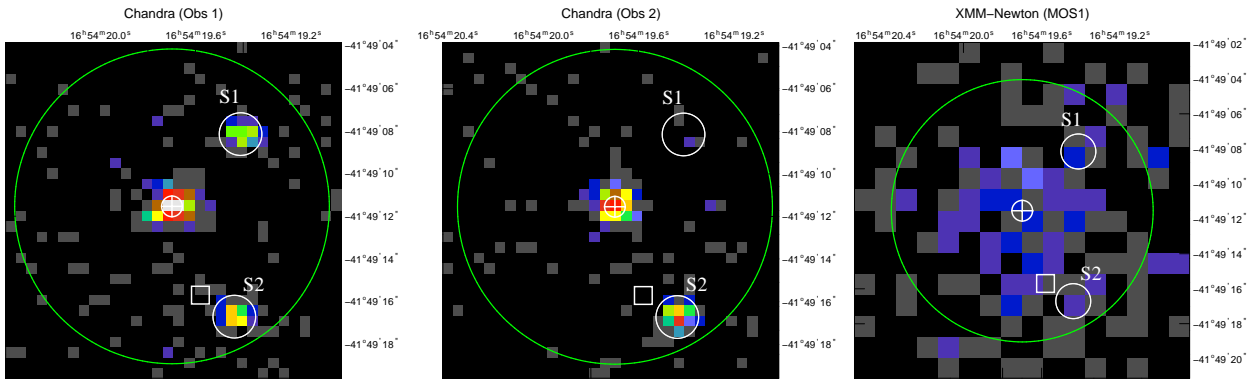

 \centering\includegraphics[width=2.11in,height=2.in]{fig5a.eps}
 \centering\includegraphics[width=2.25in,height=2.in]{fig5b.eps}
 \centering\includegraphics[width=2.1112in,height=2.in]{fig5c.eps}
\caption{
WR79 images: {\it Chandra} ACIS-S images for the two observations used
in this study with ObsID 5372 (Obs 1) and 6291 (Obs 2), and an {\it
XMM-Newton} MOS1 image (ObsID 0109490401). All images are centered at
the optical position of WR79 (SIMBAD) marked by a circle with
crosshairs. Two X-ray sources are seen in close distance from WR79:
marked by a cirle and denoted  S1 and S2. A square marks the optical
position of CCDM J16543-4149D (a binary system; SIMBAD). The green
circle has a diameter of $15\arcsec$.
}
\label{fig:wr79_images}
\end{figure*}

To quantify the variability pattern in WR46, we performed the
following exercise that is indicative of whether the
corresponding emission is variable. First, we fitted each LC with a 
constant (adopting $\chi^2$ fitting). Second, excluding the two 
`extreme' values, i.e. the maximum and minimum value, in each LC, we 
fitted the LC with a constant again. The result as a formal goodness 
of the fit is as follows. For the complete LC, the goodnes of the fit 
for a constant emission is 0.0002 (UVM2), 0.0 (UMW2), 0.49 (X-ray, pn) 
and 0.99 (X-ray, MOS). In the case with the two extreme values 
excluded, the goodness of the fit is 0.24 (UVM2), 0.0 (UMW2), 0.87 
(X-ray, pn) and 1.0 (X-ray, MOS).
We note that excluding one or two data points from a sequence of
measurements is not expected to change the global trend in a variability 
pattern of a physical quantity. Thus, we feel it is safe to conclude 
that only the UV emission of WR46 in the UVW2 filter of the optical 
monitor on-board {\it XMM-Newton} does show clear sign for {\it not} 
being constant. We believe that more X-ray and UV data, taken 
simultaneously, are needed to firmly establish whether the X-ray 
emission from WR46 is variable, and if it is, then on what timescale,
on what luminosity scale and whether it correlates with the variability
detected in other spectral domains (e.g., optical, UV).

\subsection {WR79}

The {\it XMM-Newton} observations of WR79 were analysed by
\citet{goss_11b}.
But
thanks to the {\it Chandra} superb spatial resolution, the ACIS-S
images of WR79 reveal that there are two near-by sources within 
$6\arcsec$ from WR79 (Fig.~\ref{fig:wr79_images}). We denote them as S1 
(the northern source) and S2 (the southern source).
The S1 coordinates are
$\alpha_{2000} = 16^h 54^m 19\fs41$,
$\delta_{2000} = -41\degr 49\arcmin 08\farcs15$
and WR79 is the only object in SIMBAD within $5\arcsec$ from S1
(radial distance of $4\farcs7$). Source S1 is thus unidentified yet.
The S2 coordinates are
$\alpha_{2000} =16^h 54^m 19\fs44$,
$\delta_{2000} = -41\degr 49\arcmin 16\farcs71$
(radial distance of $5\farcs9$ to WR79) and the
binary system CCDM J16543-4149D is the only object in SIMBAD within
$2\arcsec$ from S2 (radial distance of  $1\farcs9$). Source S2 is
thus unidentified yet.

It is important to note that it is possible to extract a `net' X-ray
spectrum of WR79 from the {\it Chandra} data while only a total
spectrum for all three sources, WR79, S1 and S2, can be obtained
from the {\it XMM-Newton} data (Figs.~\ref{fig:wr79_images} and
\ref{fig:wr79_spectra}). We see from Fig.~\ref{fig:wr79_images} 
that S1 is definitely a variable source and Figure~\ref{fig:wr79_spectra}
illustrates that the X-ray emission from S1 and S2 may alter
appreciably the X-ray emission from WR79 when the latter is observed
with X-ray telescope having not very high spatial resolution. 
We note that the total number of counts from sources S1 and S2 amounts to
37-47\% of that from WR79 as directly measured in the ACIS-S data of
the two {\it Chandra} observations. Also,
the X-ray emission of S1 and S2 is much softer compared to that of
WR79 and if the total spectrum of these three sources is assigned to
WR79 alone, then from analysis of such a `combined' spectrum we may draw 
conclusions about the X-ray characteristics (e.g., plasma temperature, 
X-ray absorptions, variability) that will not be quite correct.
From all this, we can safely conclude that for the moment {\it Chandra} 
observations are the only ones that can provide us with valuable 
information on the X-ray emission from the WR$+$O binary WR79.

\begin{figure*}
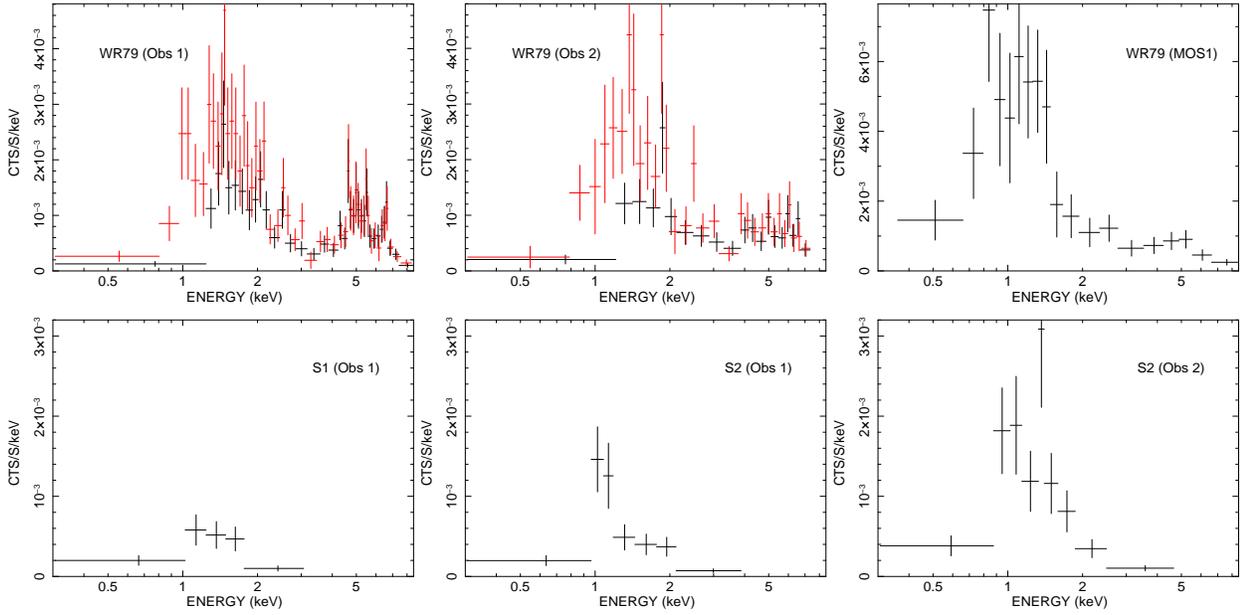

 \centering\includegraphics[width=1.58in,height=2.12in,angle=-90]
{fig6a.eps}
 \centering\includegraphics[width=1.58in,height=2.12in,angle=-90]
{fig6b.eps}
 \centering\includegraphics[width=1.58in,height=2.12in,angle=-90]
{fig6c.eps}
 \centering\includegraphics[width=1.58in,height=2.12in,angle=-90]
{fig6d.eps}
 \centering\includegraphics[width=1.58in,height=2.12in,angle=-90]
{fig6e.eps}
 \centering\includegraphics[width=1.58in,height=2.12in,angle=-90]
{fig6f.eps}

\caption{
The background-subtracted spectra of the X-ray sources from
Fig.~\ref{fig:wr79_images}.
The WR79 spectra extracted from the {\it Chandra} ACIS-S data are shown
in the first two panels of the upper row as those from the large
extraction region having a diameter of 15 arcsec (shown in green in
Fig.~\ref{fig:wr79_images}) are drawn in red while the spectra {\it
only} of WR79 itself are given in black (they are the ones used in
the current study, see Fig.~\ref{fig:spectra}). For comparison, the 
archive {\it XMM-Newton} MOS1 spectrum of WR79 is shown in the last 
panel of the upper row.
The X-ray spectra of sources S1 and S2 (Fig.~\ref{fig:wr79_images})
are shown in the lower row. No spectrum of S1 from Obs 2 is presented
since there are only 2 counts in it.
All the spectra are re-binned to have a minimum 10 cts per bin.
}
\label{fig:wr79_spectra}
\end{figure*}

\bsp

\label{lastpage}

\end{document}